\documentclass[aps,prx,twocolumn,superscriptaddress,showpacs,floatfix,longbibliography]{revtex4-1}
\usepackage{amsmath,amssymb,amsfonts,float,graphics,epsfig,epstopdf,color,verbatim,tabularx,bm,multirow,appendix,wasysym}
\usepackage[utf8]{inputenc}
\usepackage[T1]{fontenc}
\usepackage{xcolor}
\usepackage{dsfont}
\usepackage{textcomp}
\usepackage{yfonts}
\usepackage{bm}
\usepackage{amsmath}
\usepackage{subfigure}
\usepackage{mathrsfs}
\usepackage{graphicx}
\usepackage{verbatim}
\usepackage{hyperref}
\usepackage{multirow}

\usepackage{braket}
\usepackage[normalem]{ulem}
\usepackage{lipsum}
\usepackage{tikz}

\newcommand{\newSec}[1]{\textcolor{blue}{{\textit{#1}.-}}}

\begin{document}

\title{Tracking the variation of entanglement R\'enyi negativity 
: A quantum Monte Carlo study}

\author{Yi-Ming Ding}
\affiliation{Department of Physics, School of Science and Research Center for Industries of the Future, Westlake University, Hangzhou 310030,  China}
\affiliation{Institute of Natural Sciences, Westlake Institute for Advanced Study, Hangzhou 310024, China}

\author{Yin Tang}
\affiliation{Department of Physics, School of Science and Research Center for Industries of the Future, Westlake University, Hangzhou 310030,  China}
\affiliation{Institute of Natural Sciences, Westlake Institute for Advanced Study, Hangzhou 310024, China}

\author{Zhe Wang}
\affiliation{Department of Physics, School of Science and Research Center for Industries of the Future, Westlake University, Hangzhou 310030,  China}
\affiliation{Institute of Natural Sciences, Westlake Institute for Advanced Study, Hangzhou 310024, China}

\author{Zhiyan Wang}
\affiliation{State Key Laboratory of Surface Physics and Department of Physics, Fudan University, Shanghai 200438, China}
\affiliation{Department of Physics, School of Science and Research Center for Industries of the Future, Westlake University, Hangzhou 310030,  China}
\affiliation{Institute of Natural Sciences, Westlake Institute for Advanced Study, Hangzhou 310024, China}

\author{Bin-Bin Mao}
\affiliation{School of Foundational Education, University of Health and Rehabilitation Sciences, Qingdao 266000, China}

\author{Zheng Yan}
\email{zhengyan@westlake.edu.cn}
\affiliation{Department of Physics, School of Science and Research Center for Industries of the Future, Westlake University, Hangzhou 310030,  China}
\affiliation{Institute of Natural Sciences, Westlake Institute for Advanced Study, Hangzhou 310024, China}

\begin{abstract}
    Entanglement entropy has been a powerful tool for analyzing phases and criticality in pure ground states via quantum Monte Carlo (QMC). However, mixed-state entanglement, relevant to systems with dissipation, finite temperature, and disjoint regions, remains less explored due to the lack of efficient numerical methods. In this work, we present a practical and easy-to-implement QMC method within the reweight-annealing framework, enabling efficient computation of the entanglement R\'enyi negativity (RN) by tracking its variation along given parameter paths. This method is scalable, parallelizable, and well-suited for high-dimensional and large-scale simulations. Applying it to diverse scenarios—including 1D and 2D systems, ground and thermal states, and bipartite and tripartite partitions, not only the information of the underlying conformal field theory is achieved, but the role of entanglement in quantum and thermal phase transitions is revealed.
\end{abstract}
\maketitle

\newSec{Introduction}
The increasing connection between condensed matter physics and quantum information science have significantly deepened our understanding of quantum many-body behaviors. Beyond the conventional Landau-Ginzburg-Wilson framework of symmetry breaking and phase transitions~\cite{sachdev1999quantum,lifshitz2013statistical}, quantum entanglement and other intricate microcosmic degrees of freedom have become central in explaining various exotic emergent phenomena~\cite{amico2008entanglement,laflorencie2016quantum,zeng2019quantum}. 
In a bipartite composite system, the entanglement of a pure ground state can be quantified by \emph{entanglement entropy (EE)}, which captures universal information of the system including critical behaviors~\cite{KallinJS2014,Helmes2014,JRZhao2021,JRZhao2022}, continuous symmetry breaking~\cite{metlitski2011entanglement,d2020entanglement,deng2023improved,wang2024quantummontecarloalgorithm}, the underlying conformal field theory (CFT)~\cite{Calabrese2008entangle,Fradkin2006entangle,CASINI2007,JiPRR2019,Tang2020critical} and topological order~\cite{Kitaev2006,Levin2006,Isakov2011,Nussinov2006,Nussinov2009}. 
However, EE fails as a measure of entanglement in mixed states, which frequently arise in many-body physics, such as in finite-temperature Gibbs states, systems with disconnected partitions, and open quantum systems~\cite{lifshitz2013statistical,laflorencie2016quantum,zyan2018}. 

The \emph{entanglement logarithmic negativity}, referred to as \emph{negativity} in this letter, is an important entanglement monotone for mixed states that can be computed without requiring optimization procedures~\cite{vidal2002computable, plenio2005logarithmic, Audenaert2002entanglement,Eisler_2015,DeNobili_2016,BIANCHINI2016879,PhysRevB.93.115148,PhysRevB.95.165101,Shapourian_2019, Sherman2016logNegLinkCluster}. 
Given a bipartite composite system $A\cup B$ in state $\rho$, the negativity that quantifies the entanglement between $A$ and $B$ is defined as $\mathcal{E}:=\ln||\rho^{T_B}||$, where $T^B$ denotes the partial transpose operating on subsystem $B$, and $||\cdot ||$ is the trace norm, which sums over all absolute values of the eigenvalues of $\rho^{T_B}$. 
According to the positive partial transpose (PPT) criterion~\cite{peres1996separability,HORODECKI1997333,PhysRevLett.84.2726}, a nonzero value of negativity indicates the existence of entanglement. While the converse does not always hold, building on the PPT criterion, negativity has been demonstrated as a powerful tool for characterizing mixed-state entanglement~\cite{calabrese2012entanglement,Calabrese_2013_replicaTrick,calabrese2013entanglement,kulaxizi2014conformal,calabrese2014finite,de2015entanglement,wichterich2009scaling,ruggiero2016entanglement,javanmard2018sharp,lee2013entanglement,castelnovo2013negativity,wen2016topological,wen2016edge,hart2018entanglement,lu2020structure,sherman2016nonzero, Lu2020topoorder, PeterLu2024decoherence}. 
However, for general many-body systems, it is highly challenging to accurately determining the full spectrum of $\rho^{T_B}$ to compute negativity. 
Instead, the moments $\text{tr}[(\rho^{T_B})^n]$, where $n\ge 1$ is an integer, are more tractable~\cite{elben2020p3ppt,calabrese2012entanglement, Calabrese_2013_replicaTrick,calabrese2013entanglement,chuang2014negmoment_qmc,Alba2013negativity,Neven2021p3ppt}. Particularly, the \emph{R\'enyi negativity (RN)}, a R\'enyi version of negativity, has emerged as an important quantity for studying mixed-state entanglement~\cite{KHWu2020rn_prl, wybo2020rn_not_monotone, FHWang2023rn, RHFan2024quantumMemory,Alba2013negativity}. It is defined as 
\begin{equation}\label{eq:def}
    R_n(\rho) := -\ln 
    \frac{\text{tr}[(\rho^{T_{B}})^n]}{\text{tr}(\rho^n)}
\end{equation}
Although RN is not an entanglement monotone as it may increase under local operations and classical communications~\cite{wybo2020rn_not_monotone}, it can reflect many important properties of entanglement just as the negativity. For instance, it obeys the area law and the area-law coefficient is singular at the finite temperature critical point~\cite{PeterLu2019singularity, PeterLu2020structure, KHWu2020rn_prl, FHWang2023rn}. From the perspective of numerical computation, RN and negativity share a relationship analogous to that between R\'enyi EE and von Neumann EE, with the former being more tractable.

In recent years, some numerical approaches have been developed to calculate RN or the moments~\cite{Alba2013negativity, PhysRevB.97.155123, chuang2014negmoment_qmc, KHWu2020rn_prl,FHWang2023rn}. 
However, 
particularly for large-scale interacting spin/boson systems, there are few quantum Monte Carlo (QMC) works~\cite{Chung2014entanglement,wu2020entanglement}, mainly because the incremental trick commonly used in calculating R\'enyi EE~\cite{hastings2010measuring,humeniuk2012quantum,d2020entanglement} becomes much more complex and difficult in the case of RN, since the lowest order of nontrivial RN is third and the boundary condition of imaginary time needs to be incrementally changed from a glued space-time manifold to a twistedly glued one. 
Therefore, developing an efficient QMC algorithm that is both easy to implement and capable of accurately computing RN is a crucial yet challenging task. 

Furthermore, the absence of such algorithms has left many important questions about mixed-state entanglement unexplored. 
For example, one might naturally assume that the infinite quantum correlation length at a quantum critical point necessarily implies long-range entanglement. However, this is not always the case. Quantum correlation length does not directly measure entanglement: for instance, gapped topological phases can exhibit long-range entanglement despite having finite correlation lengths.
Meanwhile, the classical Ising model has no entanglement even at the thermal critical point with infinite correlation length. Nevertheless, the thermal phase transition of a quantum system has both thermal and quantum fluctuations. Due to the existence of quantum fluctuation, the $T_c$ of thermal phase transition of a quantum model is usually lower than a pure classical model. It can thus be seen that quantum fluctuations are involved in the thermal phase transition of quantum systems. A fundamental question is what role the entanglement plays in this case.

In this letter, we have filled the technical gap by presenting an efficient QMC algorithm within the \emph{reweight-annealing frame}~\cite{YMDing2024reweight,ZheWang2024reweight,jiang2024fermion,ym2025magic} and explored several fundamental questions about the entanglement in mixed states.

\begin{figure}[ht!]\centering 
    \includegraphics[width=0.5\textwidth]{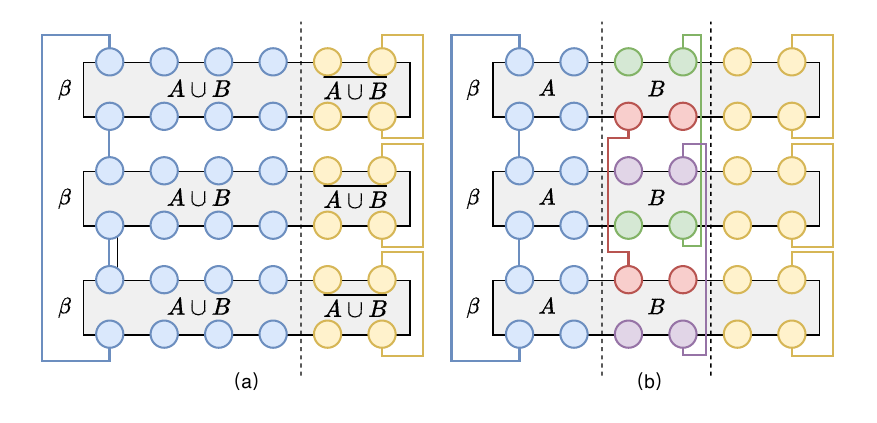}
    \caption{
        The horizontal axis represents spatial spins (depicted as circles), while the vertical axis corresponds to imaginary time. Spins of the same color are connected (or glued) along the imaginary time direction by straight lines of the same color.
        (a) Manifold for $Z_3$: The three replicas are sequentially connected along the imaginary time direction for the system $A\cup B$. The environment, denoted by $\overline{A\cup B}$, is traced out by imposing periodic boundary conditions on each replica.
        (b) manifold for $Z_3^{T_B}$: The connections among three replicas in subsystem $A$ (the blue part) and the environment $\overline{A\cup B}$ (the yellow part) remain identical to that in the left manifold. However, in subsystem $B$, the connections are not in sequence due to the operation $T_B$.
    }\label{fig:manifold}
\end{figure}
\newSec{Basic idea} 
For convenience, we define two generalized partition functions: $Z_n:=\text{tr}(\rho^n)$ and $Z_n^{T_B}:=\text{tr}[(\rho^{T_B})^n]$, which correspond to two types of manifolds (illustrated by $n=3$ in Fig.~\ref{fig:manifold}) in the path integral formulation. Thus ${R}_n=-\ln Z_n^{T_B}/Z_n$. Here, we define $\rho:=e^{-\beta H}$, which is not normalized in the context of QMC.
Typically, $Z_n^{T_B}$, $Z_n$ and ${R}_n$ are functions of some real parameter $\lambda>0$ (such as temperature or a Hamiltonian parameter). Our starting point is to consider the difference of $R_n$ at two different parameter points $\lambda'$ and $\lambda''$ ($\lambda'<\lambda''$), such that
\begin{equation}\label{eq:key_eq}
    \begin{split}
        {R}_n(\lambda'')-{R}_n(\lambda') 
        =& \ln\frac{Z_n^{T_B}(\lambda')}{Z_n^{T_B}(\lambda'')} - \ln\frac{Z_n(\lambda')}{Z_n(\lambda'')}.
    \end{split}
\end{equation}
The ratio $[Z_n^{T_B}(\lambda')/Z^{T_B}_n(\lambda'')]$ in Eq.~(\ref{eq:key_eq}) can be further interpreted as a \emph{reweighting operation}~\cite{ferrenberg1988reweighting, ferrenberg1989reweighting, Troyer2004reweighting,xuxiaoyan}.
If express $Z_n^{T_B}(\lambda)=\sum_iW_i(\lambda)$, where $\{W_i(\lambda)\}$ are the weights, then the ratio can be viewed as the average of weight ratios, since
$Z_n^{T_B}(\lambda')/Z^{T_B}_n(\lambda'')=\langle W(\lambda')/W(\lambda'')\rangle_{Z_n^{T_B}(\lambda'')}$, where  $W(\lambda)\in \{W_i(\lambda)\}$ and the expectation is computed with respect to samples from $Z_n^{T_B}(\lambda'')$.
The specific form of the estimator $\langle W(\lambda')/W(\lambda'')\rangle_{Z_n^{T_B}(\lambda'')}$ depends on the QMC method we use, which transforms the quantum degrees of freedom in $Z_n^{T_B}(\lambda')$ to some classical ones.
For illustration, we use the stochastic series expansion (SSE) method in this letter~\cite{sandvik1998stochastic, sandvik2003stochastic, yan2019sweeping, yan2020improved}, and the corresponding estimators can be found in the Supplemental Material~\cite{supmat} (see also references~\cite{sandvik1998stochastic, sandvik2003stochastic,yan2019sweeping,yan2020improved,Zhou2022,zhou2020,yan2023quantum,Calabrese_2013_replicaTrick,laflorencie2016quantum} therein).
Similarly, we can apply the same reasoning to $[Z_n(\lambda')/Z_n(\lambda'')]$. 
It is important to emphasize that the view of reweighting is crucial, as it allows us to estimate the difference $[{R}_n(\lambda'')-{R}_n(\lambda')]$ from Eq.~(\ref{eq:key_eq}) by estimating $[Z_n^{T_B}(\lambda')/Z^{T_B}_n(\lambda'')]$ and $[Z_n(\lambda')/Z_n(\lambda'')]$ in QMC simulations. 
Additionally, to avoid overflow/underflow issues, we have to store their logarithms in computers.

The procedure described above is sufficient to derive the relative values of $R_n(\lambda)$ at different $\lambda''$ when $\lambda'$ is fixed at $\lambda_0$, called a \emph{reference point}. This allows for the study of various behaviors, such as locating extreme values and singularities of $R_n(\lambda)$. Moreover, if ${R}_n(\lambda_0)$ is known, we can compute the exact values of ${R}_n(\lambda)$. 
Typically, finding such a reference point is straightforward. 
For example, when $\lambda=\beta\equiv 1/T$, the inverse temperature, we can set $\beta_0=0$, which corresponds to a completely disordered system where both classical and quantum correlations are destroyed, giving ${R}_n(\beta_0)=0$.
In practice, we use a sufficiently small value (e.g. $\beta_0=10^{-8}$) in our simulations, and the adequacy of the choice can be verified by considering a smaller one to check the convergence. 
Similarly, when $\lambda$ represents the coupling strength of interacting terms in a Hamiltonian related to the entangled boundary, setting $\lambda_0=0$ would make the two subsystems completely independent, resulting in ${R}_n(\lambda_0)=0$ again.
Therefore, depending on our choice of $\lambda$, both finite and zero temperature properties of $R_n$ can be explored.
For illustrative purpose, we now set $\lambda=\beta$, and other choices of $\lambda$ can be similarly discussed. 
By identifying $\lambda_0\equiv \beta_0\equiv 0$, Eq.~(\ref{eq:key_eq}) can be reformulated as 
$    {R}_n(\beta)=\ln\frac{Z_n^{T_B}(0)}{Z_n^{T_B}(\beta)} - \ln \frac{Z_n(0)}{Z_n(\beta)}$.

\newSec{Importance sampling and annealing scheme} 
Although the ideas of reweighting and reference point are simple, simulations become inefficient if $\beta$ is far away from $\beta_0$ in the parameter space, due to the significant differences between the distributions $Z_n^{T_B}(\beta)$ and $Z_n^{T_B}(\beta_0)$,  requiring an excessive number of samples. To address this, we divide $[\beta_0,\beta]$ into $m$ small subintervals, i.e.
 $   \ln\frac{Z_n^{T_B}(0)}{Z_n^{T_B}(\beta)} = \sum_{k=1}^{m}\ln \frac{Z_n^{T_B}(\beta_{k-1})}{Z^{T_B}_n(\beta_k)} $ and
    $\ln \frac{Z_n(0)}{Z_n(\beta)} = \sum_{k=1}^m \ln \frac{Z_n(\beta_{k-1})}{Z_n(\beta_k)}$,
where $\beta_m\equiv \beta$. 
Then, efficient estimation is ensured when adjacent $\beta_k$ values are close to satisfy the importance sampling~\cite{neal2001annealed}.
In addition, the method's effectiveness depends on how $[\beta_0,\beta]$ is divided, as the computational cost also scales with the number of subintervals. Simple approaches, such as equal divisions or geometric progressions with large/small ratios, are inefficient since the required number of subinterval density varies across parameter regions and different system sizes.

To address the problem, we adopt the \emph{annealing scheme}, which requires a polynomial number of subintervals and has demonstrated excellent performance in practical partition function calculations~\cite{YMDing2024reweight}.
The core idea is to approximately restrict $[Z_n^{T_B}(\beta_{k-1})/Z^{T_B}_n(\beta_{k})]$ and $[Z_n(\beta_{k-1})/Z_n(\beta_{k})]$ to be some constant $\epsilon\lesssim 1$ for all sizes and $\beta_k$, to ensure the importance sampling. 
With some calculations, we find that the total number of subintervals $F(L)$ required scales as $F(L)\approx (\beta-\beta_0)\Lambda nL^d/|\ln\epsilon|$, where $d$ is the system dimension, $L$ is the length of system, and $\Lambda$ is a constant related to the energy density. 
Notably, an incorrect choice of $\Lambda$ just effectively changes the value of $\epsilon$, making $\epsilon$ the sole hyperparameter in the algorithm. Similarly, one can discuss when $\lambda$ is some other parameter in the Hamiltonian, and the division number also scales polynomially with the system size. Further details, including the derivation of $F(L)$ can be found in the Supplemental Material~\cite{supmat}.

\newSec{Ground states of tripartite 1D antiferromagnetic dimerzied Heisenberg model (ADHM)}
\begin{figure}[htbp]\centering 
    \includegraphics[width=0.45\textwidth]{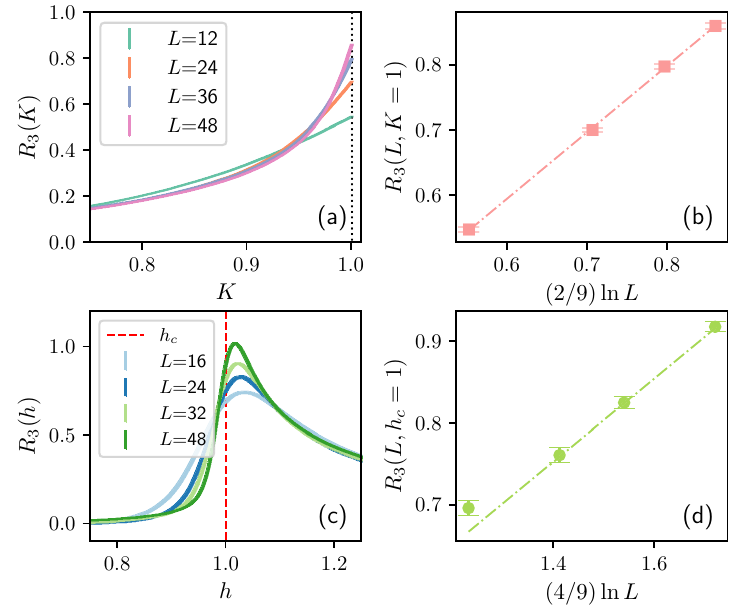}
    \caption{
        With $\beta=L$ to extrapolate to the ground state: (a) The variation of $R_3(K)$ with respect to the parameter $K$ for the tripartite (1+1)D ADHM ring;
        (b) The central charge $c=1.02(2)$ of the underlying CFT for the Heisenberg model is fitted;
        (c) The variation of $R_3(h)$ with respect to the parameter $h$ for the (1+1)D transverse field Ising model (TFIM) under half bipartition;
        (d) The central charge of the underlying CFT at the critical point is fitted to be $c=0.51(3)$. Notably, the data of $L=16$, which exhibits significant finite-size effects and clearly deviates from the trend, is excluded from the fitting.
    }\label{fig:1d_rn}
\end{figure}
For the first example, we consider a 1D ring of ADHM whose Hamiltonian is $H_{\text{ADHM}}=\sum_{i\in\text{odd}}\mathbf{S}_i\cdot\mathbf{S}_{i+1}+K\sum_{i\in\text{even}}\mathbf{S}_i\cdot\mathbf{S}_{i+1}$. 
We partition the system into three connected and equally sized regions: subsystems $A$, $B$, and the environment $\overline{A\cup B}$, with boundary bonds having a coupling strength of $K$. When $K=0$, the three regions $A$, $B$, and $\overline{A\cup B}$ are completely independent, thus we take $K_0=0$ as the reference point and gradually increase $K$ from $K_0$ to $K=1$ in simulations.
By tracing out the environment $\overline{A\cup B}$ according to Fig.~\ref{fig:manifold}, the entanglement between subsystem $A$ and $B$ is in the context of mixed state. 
As shown in Fig.~\ref{fig:1d_rn}(a), the entanglement quantified by $R_3$ increases monotonically when increasing $K$. 
This indicates that the entanglement between $A$ and $B$ arises from the Heisenberg interaction at the boundaries.
Moreover, at $K=1$, the model is described by a $(1+1)$D CFT, and $R_n$ satisfies~\cite{calabrese2012entanglement,Calabrese_2013_replicaTrick} 
\begin{equation}
    R_{n}=
    \begin{cases}
    \frac{c}{12}(n-\frac{1}{n})\ln\bigg(
        \frac{L}{\pi a} \frac{\sin{\frac{\pi l_1}{L}} \sin{\frac{\pi l_2}{L}}}{\sin{\frac{\pi (l_1+l_2)}{L}}}  \bigg) + \mathcal{O}(1),  n\in\text{odd} \\
        \frac{c}{6}(\frac{n}{2}-\frac{2}{n})\ln\bigg(
        \frac{L}{\pi a} \frac{\sin{\frac{\pi l_1}{L}} \sin{\frac{\pi l_2}{L}}}{\sin{\frac{\pi (l_1+l_2)}{L}}}  \bigg) + \mathcal{O}(1),  n\in\text{even}
    \end{cases}
\end{equation}
where $c$ is the central charge, $a$ is the lattice constant, $L$ is total length of the ring, and $l_1$ and $l_2$ are the lengths of subsystems $A$ and $B$, respectively. Note that the standard negativity $\mathcal{E}$ can be obtained through taking $n\to 1$ for even $n$. For $n=3$, we therefore have $R_3\propto (2c/9)\ln L$. As shown in Fig.~\ref{fig:1d_rn}(b), by fitting $R_3(K=1)$ with the data, we successfully extract $c=1.02(2)$, which is consistent with the theoretical value $1$~\cite{blote1986cft}. 

\newSec{Ground states of bipartite 1D TFIM}
We next consider the TFIM, and the Hamiltonian is given by $H=-\sum_{\langle ij\rangle}Z_iZ_j -h\sum_{i}X_i$, where $\langle ij\rangle$ denotes the nearest neighbor. 
For the 1D ring, we consider half bipartition without the environment.
By fixing $J=1$, the quantum critical point (QCP) is at $h_c=1$, which is also described by a CFT with central charge $c_{\text{Ising}}=1/2$. We take $h_0=0$ to be the reference point, and Fig.~\ref{fig:1d_rn}(c) shows that $R_3(h)$ is maximized at the QCP. When increasing the system length $L$, it is observed that $R_3(L,h_c)$ diverges. 
In addition, in the case of half bipartition, $R_n$ satisfies
\begin{equation}\label{eq:tfim_cft}
    R_{n}(h_c)=
    \begin{cases}
    \frac{c}{6}(n-\frac{1}{n})\ln\bigg(
        \frac{L}{\pi a}\sin\frac{\pi l}{L}  \bigg) + \mathcal{O}(1), & n\in\text{odd} \\
        \frac{c}{3}(\frac{n}{2}-\frac{2}{n})\ln\bigg(
        \frac{L}{\pi a}\sin\frac{\pi l}{L}  \bigg) + \mathcal{O}(1), & n\in\text{even}
    \end{cases}
\end{equation}
for a (1+1)D CFT, where $l$ is the length of subsystems $A$ and $B$. With Eq.~(\ref{eq:tfim_cft}), the central charge is obtained to be $c=0.51(3)$, which also accords with the theoretical value.
In both the tripartite 1D ADHM and the bipartite 1D TFIM, the logarithmic corrections to $R_n$ at the CFT points deviate from the expected relation $R(2l) = 2R(l)$, signaling the presence of long-range entanglement~\cite{Lu2020topoorder,PeterLu2020structure,PeterLu2024decoherence}. In contrast, a system with only short-range entanglement would yield an extensive $R_n$, with no subleading terms beyond the area law.
\begin{figure}[htbp]\centering 
    \includegraphics[width=0.45\textwidth]{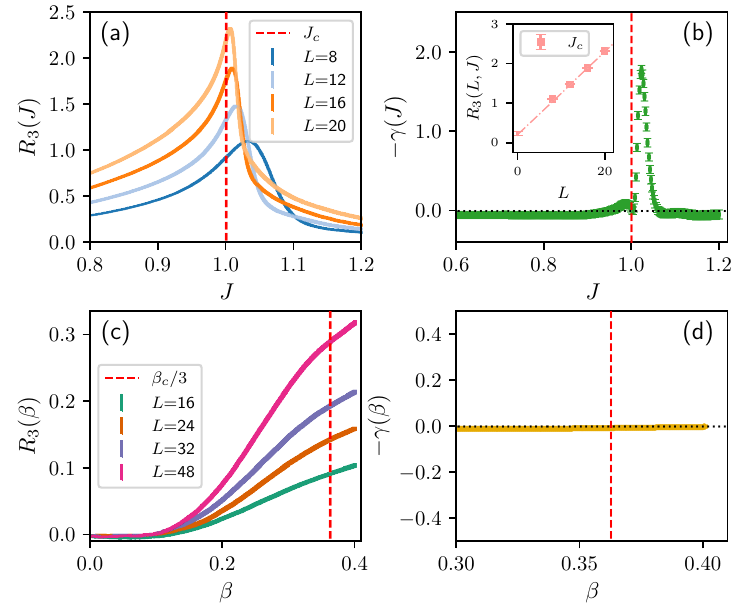}
    \caption{
        For the 2D TFIM on the square lattice: 
        (a) fixing $h=3.04438$, the variation of $R_3(J)$ as a function of $J$, where $\beta=L$ to extrapolate to the ground state;
        (b) the area law correction $-\gamma(J)$ of $R_3(J)$ for $J\in[0,1.2]$. The inset panel shows the fitting of area law with the maximum value of $R_3(L,J)$ for each $L$, and the area law coefficient is $\gamma=-0.225(49)$; 
        (c) fixing $J=1$ and $h=2.75$, the variation of $R_3(\beta)$ as a function of $\beta$. 
        (d) the area law correction $-\gamma(\beta)$ of $R_3(\beta)$ for $\beta\in(0,0.4]$, and the inset shows the fitting of the area law at $\beta=\beta_c/3$, where $-\gamma(\beta_c)=0.002(7)$;
    }\label{fig:2d_rn}
\end{figure}

\newSec{Ground states and Gibbs states of bipartite 2D TFIM}
We similarly investigated $R_3$ of the 2D TFIM on a $L\times L$ square lattice with periodic boundary condition, considering the cornerless half-bipartition. The QCP of its ground state occurs at $h/J= 3.04438(2)$~\cite{dyj2022qmc}. 
For convenience, we fix $h=h_c\equiv 3.04438$ and tune $J$, so that the QCP is positioned at $J_c\approx 1$. 
To study the ground-state entanglement, we consider $R_3$ as a function of $J$ and take $J_0=0$ as the reference point. 
Furthermore, when $h/J<3.04438$, increasing the temperature can induce a thermal phase transition from a ferromagnetic to a paramagnetic phase~\cite{Hesselmann2016isingTCP}. Thus we can set a large $h$ to highlight the quantum fluctuations, and study the finite-temperature mixed-state entanglement across a thermal critical point (TCP). We choose $J=1$ and $h=2.75$, and the TCP located at $\beta_c=1/T_c=1.0874(1)$~\cite{KHWu2020rn_prl}. 

For the ground states, Fig.~\ref{fig:2d_rn}(a) illustrates the behavior of $R_3(J)$ as a function of $J$ for different system sizes. Similar to the 1D TFIM, $R_3(J)$ reaches a maximum and diverges at the QCP, indicating that the QCP exhibits stronger quantum correlations compared to the gapped phases. In the gapped phases, no long-range entanglement is expected, and we should have $R_3\sim aL -\gamma$ with $\gamma=0$. 
Near the critical point, we observe that the area law breaks down, with $R_3$ for small system sizes exceeding that of larger sizes, as shown in Fig.~\ref{fig:2d_rn}(a). However, as the system moves further from the critical point, the area law is restored.
If we forcibly fit $R_3$ to the area law, the constant term $\gamma$ fluctuates significantly near the critical point, as shown in Fig.~\ref{fig:2d_rn}(b). In other regions, it remains zero within error bars.
In principle, $R_n$ can be understood as the difference of free energies between two space-time manifolds with different boundary conditions. 
Since the only distinction lies at the boundary, it should follow an area law. The deviation from the area law in our numerical data is due to finite-size effects in both space and time, as the gaplessness related to the phase transition requires larger system sizes. Further details on finite-size effects are provided in the Supplemental Material~\cite{supmat}.

To reduce finite-size effects, we use the peaks of $R_3(J)$ across different system sizes to fit the area law at the QCP, as shown in the inset of Fig.~\ref{fig:2d_rn}(b)\footnote{For large sizes, this result converges to that obtained by fitting at a fixed parameter point.}.
Its constant correction extracted from the fit is $\gamma =-0.225(49)$~\footnote{Notice we use $R_3\sim aL -\gamma$ to fit, thus the final constant correction implies a positive sub-leading correction.}, which signifies a breakdown of the local behavior of RN and reflects the long-range entanglement at the QCP here. This constant correction is consistent with the theoretical prediction~\cite{metlitski2009entanglement}.
In addition, RN inherently removes the contribution of ground state degeneracy due to the denominator in its definition, whereas EE captures it in the correction term as $\gamma=-\ln(N_\text{deg})$, where $N_\text{deg}$ denotes the ground state degeneracy~\cite{PhysRevB.80.184421}. In this example of ferromagnetic phase, we have $\gamma=0$ while $N_\text{deg}=2$. This reflects the advantage of RN in computing mixed-state properties, where it naturally filters out degeneracy effects.

For the Gibbs states, Fig.~\ref{fig:2d_rn}(c) shows that $R_3(\beta)$ increases monotonically from the infinite high temperature $\beta_0$ to $\beta=0.4$.
Notably, the three replicas effectively scale the inverse temperature by a factor of $1/3$ in partition function $Z_3$, and the related critical properties are shifted to $\beta_c/3$~\footnote{The term $\mathrm{tr} (\rho^3)\propto \text{tr}(e^{-3\beta H})$ will lead a singularity at $\beta_c/3$}. This has been confirmed in the calculation of the divergence of the first derivative of $a$, the area-law coefficient of $R_3$, in the quantum spherical model and 2D TFIM~\cite{KHWu2020rn_prl}. 
As seen in Fig.~\ref{fig:2d_rn}(c), at the critical point $\beta_c/3$, no significant behavior associated with the TCP is observed for $R_3$ in our calculations. 
Furthermore, we find that $R_3$ perfectly obeys the area law and $\gamma\sim0$ for all the $\beta$ we computed, as shown in Fig.~\ref{fig:2d_rn}(d). 
Despite choosing $h = 2.75$, close to the QCP and dominated by strong quantum fluctuations, our findings indicate that thermal phase transitions remain short-range entangled, highlighting a key distinction from the long-range entanglement inherent in quantum criticality.
Moreover, the increase in $R_3$ when decreasing the temperature reflects an enhancement of quantum properties. 
However, the quantum correlations remain short range ($\gamma\sim 0$), indicating that the phases beyond or below the TCP are both trivial in terms of quantum correlation. 

In addition, we discuss the derivatives of $R_n$ with respect to some parameter $\lambda$. Given that $R_n(\lambda)=-\ln Z_n^{T_B}(\lambda)+\ln Z_n(\lambda)$, it follows that if the $k$th-order derivatives of both effective free energy terms are non-singular at $\lambda$, then $d^kR_n(\lambda)/d\lambda^k$ must also remain non-singular. Moreover, since the difference between the two systems associated with $Z_n^{T_B}$ and $Z_n$ arises solely from their boundary conditions, their critical points $\lambda_c$ should remain identical in the thermodynamic limit. Consequently, any singularity in a higher-order derivative of $R_n$ must occur precisely at $\lambda_c$.

\newSec{Summary and discussions}
We have introduced an efficient algorithm for calculating R\'enyi negativity in quantum Monte Carlo simulations, built on the reweight-annealing framework. This algorithm is easy to implement, has polynomial computational complexity, and is naturally parallelizable. By enabling the calculation of R\'enyi negativity along a full parameter trajectory in parameter space, our method proves highly versatile. It is not only valuable for investigating entanglement behavior across different phases and near critical points, but also for fitting correction terms in area laws to extract universal quantities such as central charge and topological order in quantum many-body systems. We have validated the approach in various scenarios, including 1D and 2D systems, ground and Gibbs states, as well as bipartition and tripartition setups. 
Our results of the R\'enyi negativity further reveals a detailed understanding of the intrinsic mechanisms governing thermal and quantum critical points in many-body systems.
An intriguing direction for future research is to investigate the role of entanglement in strong-to-weak spontaneous symmetry breaking in mixed quantum states~\cite{lessa2025strong,lessa2025mixed}.

\newSec{Acknowledgements}
The authors sincerely thank Tsung-Cheng Lu for kindness and valuable discussions. The authors also thank Fo-Hong Wang, Wei Zhu, Rui-Zhen Huang, and Ying-Jer Kao for their helpful discussions. The work is supported by the Scientific Research Project (No.WU2024B027) and the Start-up Funding of Westlake University. ZW is supported by the China Postdoctoral Science Foundation under Grants No.2024M752898. 
BBM acknowledge the Natural Science Foundation of Shandong Province, China (Grant No. ZR2024QA194) and NSFC Grant No. 12247101.
The authors thank the high-performance computing center of Westlake University and the Beijng PARATERA Tech Co.,Ltd. for providing HPC resources.

\bibliography{ref}

\onecolumngrid
\setcounter{equation}{0}
\setcounter{figure}{0}
\section{Estimators in reweighting operations}
We focus the stochastic series expansion (SSE) method~\cite{sandvik1998stochastic, sandvik2003stochastic,yan2019sweeping,yan2020improved}, and other quantum Monte Carlo (QMC) methods can be similarly discussed in principle.
If $\lambda=\beta$, the estimators for $[Z_n^{T_B}(\beta_{k-1})/Z^{T_B}_n(\beta_k)]$ and $[Z_n(\beta_{k-1})/Z_n(\beta_k)]$ are universal as long as the model is sign-problem free.
For a given Hamiltonian $H$ and $\rho=e^{-\beta H}$, we have 
\begin{equation}\label{eq:deduction}
    \begin{split}
        Z_n^{T_B}(\beta_{k-1}) 
        = &\text{tr}(\{
            \rho^{T_B} \}^n )\\
            = &\text{tr}
            \Bigg( \bigg\{ \bigg[
                \sum_{s=0}^{\infty}\frac{\beta^s_{k-1}}{s!}(-H)^s
                \bigg]^{T_B} 
            \bigg\}^n \Bigg)   \\
        = &\text{tr}
        \Bigg( \bigg\{ \bigg[
                \sum_{s=0}^{\infty}\bigg(\frac{\beta_{k-1}}{\beta_{k}}\bigg)^s\frac{\beta^s_{k}}{s!}(-H)^s
                \bigg]^{T_B} 
            \bigg\}^n \Bigg),   \\
    \end{split}
\end{equation}
where $s$ is the order of Taylor expansion, i.e., the total number of non-identity operators of the SSE configuration. Therefore, we have  
\begin{equation}
    \frac{Z_n^{T_B}(\beta_{k-1})}{Z_n^{T_B}(\beta_k)}= \Bigg\langle  \bigg(\frac{\beta_{k-1}}{\beta_{k}}\bigg)^{s}   \Bigg\rangle_{Z_n^{T_B}(\beta_k)},
\end{equation}
where $\langle ...\rangle_{Z_n^{T_B}(\beta_k)}$ means that the sampling is performed on $Z_n^{T_B}(\beta_k)$. Similarly, 
\begin{equation}
    \frac{Z_n(\beta_{k-1})}{Z_n(\beta_k)}= \Bigg\langle  \bigg(\frac{\beta_{k-1}}{\beta_{k}}\bigg)^{s}   \Bigg\rangle_{Z_n({\beta_k})}
\end{equation}
If $\lambda$ is a non-temperature parameter, the estimators typically depend on the model. 

For example, if $\lambda=h$ in the Hamiltonian 
\begin{equation}
    H=-\sum_{\langle ij\rangle}Z_iZ_j -h\sum_{i}X_i
\end{equation}
of the transverse field Ising model (TFIM), we have 
\begin{align}
    \frac{Z_n^{T_B}(h_{k-1})}{Z_n^{T_B}(h_k)}= \Bigg\langle  \bigg(\frac{h_{k-1}}{h_{k}}\bigg)^{s_{h}}   \Bigg\rangle_{Z_n^{T_B}(h_k)} \label{eq:h_0}
    \\
    \frac{Z_n(h_{k-1})}{Z_n(h_k)}= \Bigg\langle  \bigg(\frac{h_{k-1}}{h_{k}}\bigg)^{s_{h}}   \Bigg\rangle_{Z_n({h_k})}\label{eq:h_1}
\end{align}
where $s_{h}$ is the number of operators related to the transverse field $h$ only in the series expansions no matter in the manifold $Z_n^{T_B}(h_k)$ or $Z_n(h_k)$. 
The details of the SSE algorithm for TFIM can be found in Ref.~\cite{sandvik2003stochastic,Zhou2022,zhou2020,yan2023quantum}.

\section{Annealing scheme ($\lambda=\beta$)}
The target of the annealing scheme is to provide an economic way to divide $[\beta_0,\beta]$ into $m$ subintervals $\{[\beta_{k-1},\beta_k]\}$ to ensure the importance sampling, such that we can compute $[Z_n^{T_B}(\beta_0)/Z_n^{T_B}(\beta)]$ and $[Z_n(\beta_0)/Z_n(\beta)]$ via 
\begin{equation}
    \begin{split}
        \frac{Z_n^{T_B}(0)}{Z_n^{T_B}(\beta)} &= \prod_{k=1}^{m}\frac{Z_n^{T_B}(\beta_{k-1})}{Z^{T_B}_n(\beta_k)} 
        \\
        \frac{Z_n(0)}{Z_n(\beta)} &= \prod_{k=1}^{m}\frac{Z_n(\beta_{k-1})}{Z_n(\beta_k)}
    \end{split}
\end{equation}
Remember that we have choose the reference point $\beta_0\equiv 0$. We mainly focus on $\lambda=\beta$ in this material, and it is similar to discuss other choices of $\lambda$. The slight difference will be introduced in the end of the next section. 

For a vanilla partition function $Z(\beta)=\text{tr}(e^{-\beta H})$ of some quantum spin model, each classical configuration sampled in SSE simulations is composite of both the states of spins and operators, thus the number of non-identity operators $s_{\text{vn}}$ is exactly some observable in this language, which has the relation  $E(\beta)=-\langle s_{\text{vn}}\rangle_{Z(\beta)}/\beta$ with the energy $E(\beta)$. In other words, $\langle s_{\text{vn}}\rangle_{Z(\beta)}\sim \beta |E(\beta)|$.

Suppose $\bar{s}_{\text{tot}}$ and ${s}_{\text{tot}}$ are the corresponding numbers of non-identity operators of $Z^{T_B}_n(\beta)$ and $Z_n(\beta)$. Since $Z_n(\beta)=Z(n\beta)$, we directly obtain $\langle s_{\text{tot}}\rangle_{Z_n(\beta)}=\langle s_{\text{vn}}\rangle_{Z(n\beta)}\sim n\beta|E(\beta)|$. 
Next, consider $\bar s_{\text{tot}}$. In fact, the difference between $\langle \bar{s}_{\text{tot}}\rangle_{Z_n^{T_B}(\beta')}$ and $\langle s_{\text{tot}}\rangle_{Z_n(\beta')}$, where $\beta'\in[\beta_0,\beta]$, reflects the existence of entanglement because
\begin{equation}
    \begin{split}
        R_n(\beta)=&\ln\frac{Z_n^{T_B}(0)}{Z_n^{T_B}(\beta)} - \ln \frac{Z_n(0)}{Z_n(\beta)}\\
        =&-\int_{0}^{\beta}d\beta' \bigg[ \frac{d\ln Z_n^{T_B}(\beta')}{d\beta'}
        -  \frac{d\ln Z_n(\beta')}{d\beta'}\bigg]\\
        =&-\int_0^{\beta}d\beta'\bigg[ \frac{1}{Z_n^{T_B}(\beta')}\frac{d Z_n^{T_B}(\beta')}{d\beta'} -
         \frac{1}{Z_n(\beta')}\frac{d Z_n(\beta')}{d\beta'}\bigg] \\
        =& -\int_0^{\beta}d\beta' \frac{1}{\beta'}\bigg[
            \langle \bar s_{\text{tot}}\rangle_{Z_n^{T_B}(\beta')} - \langle s_{\text{tot}}\rangle_{Z_n(\beta')}\bigg]\\
    \end{split}
\end{equation}
Since the amount of entanglement is an extensive quantity, therefore we should have $\bar{s}_{\text{tot}}\sim n\beta|E(\beta)|$ satisfies as well. With these knowledge, we can now introduce the annealing scheme. We take $[Z_n^{T_B}(\beta_{k-1})/Z_n^{T_B}(\beta_k)]$ for illustrations, and it is similar to discuss $[Z_n(\beta_{k-1})/Z(\beta_k)]$. For brevity, from now, we drop the subscript of $\langle\cdots  \rangle$, which should be the corresponding $Z^{T_B}_n(\beta_k)$, and write $\bar s_{\text{tot}}\equiv s$.

We first require $[Z_n^{T_B}(\beta_{k-1})/Z_n^{T_B}(\beta_k)]=\epsilon<1$ for all $k$, where $\epsilon$ is some constant close to 1 to ensure the importance sampling when estimating $[Z_n^{T_B}(\beta_{k-1})/Z_n^{T_B}(\beta_k)]$ with $\langle(\beta_{k-1}/\beta_k)^{s} \rangle$ at $\beta_k$. If $\epsilon=1$, the two distributions must be exactly the same and this is some trivial reweighting operation. 
As we mentioned above, $s$ is some observable, therefore we can achieve some estimation $s^*$ for it from simulations (same as we measure the energy) at $\beta_k$. Hence, we can bring such an $s^*$ into 
\begin{equation}\label{sm:eq:est}
    \epsilon = \bigg\langle \bigg(\frac{\beta_{k-1}}{\beta_k}\bigg)^{s}\bigg\rangle \approx \bigg(\frac{\beta_{k-1}}{\beta_k}\bigg)^{s^*}.
\end{equation}
Eq.~(\ref{sm:eq:est}) enables us to determine the position of $\beta_{k-1}$ only with $\beta_{k}$ since 
\begin{equation}\label{sm:eq:ddd}
    \beta_{k-1} = \epsilon^{1/s^*}\beta_k 
\end{equation} 
Therefore, if we start from $\beta_m$, using Eq.~(\ref{sm:eq:est}), we can determine the rest of $\beta_{m-1},\cdots,\beta_1$, and the procedure stops when we encounter some $\beta_j$ such that $\beta_{j-1}<\beta_0$. Each determination of $\beta_{k-1}$ relys on some simulations on $\beta_k$, and in this sense, the inverse temperature gradually goes from $\beta_m$ to $\beta_0$. This is akin to a standard quantum annealing or thermal annealing process where we tune some parameter, so we call this way of dividing $[\beta_0,\beta]$ an annealing scheme.  

To enable parallelizations, we can manually set the value of $s^*$ rather than estimating it from simulations. As energy $E\sim L^d$, where $L$ is the system length and $d$ is the dimension, we can set $s^*\simeq \Lambda_k n\beta_k L^d$, where $\Lambda_k$ is another constant factor related to the energy density.
Then, Eq.~(\ref{sm:eq:ddd}) reduces to 
\begin{equation}
    \beta_{k-1}=\epsilon^{1/(\Lambda_kn\beta_kL^d)}\beta_k 
\end{equation}
If we manually set the value of $\Lambda_k$ with some incorrect and uniform $\Lambda$, this would effectively change the true value of  $[Z_n^{T_B}(\beta_{k-1})/Z_n^{T_B}(\beta_k)]$ from $\epsilon$, what we set at the beginning, to some other $\epsilon_k$. Therefore the dependence on $k$ can be absorbed into the base, i.e. 
\begin{equation}\label{sm:eq:ppp}
    \beta_{k-1}=\epsilon_k^{1/(\Lambda n\beta_kL^d)}\beta_k 
\end{equation}
If one preknows the variation of $\Lambda_k(\beta)$ or $\epsilon_k$ as a function of $\beta$, everything would be basically same with that in Eq.~(\ref{sm:eq:ddd}), but this is not a general case. On another hand, if we further simplify Eq.~(\ref{sm:eq:ppp}) by setting  $\epsilon_k\equiv \epsilon$ for all $k$, this operation would just make the value of $[Z_n^{T_B}(\beta_{k-1})/Z_n^{T_B}(\beta_k)]$ achieved in simulations varies with different $k$, but usually at the same order of magnitude if no critical point lies within $[\beta_0,\beta]$. 
Such a critical point may change the value of $\Lambda_k$ sharply near that point. 
However, our method is not for locating the critical point, which can be obtained by considering some order parameter or dimensionless quantity such as the Binder cumulant. Therefore, we can assume that we know the position of the critical point, and on the two sides of that point, we can adopt different $\epsilon$ to avoid the problem.

To sum up, by manually setting some $\Lambda$, we only have one hyperparameter $\epsilon$ to determine. 
Smaller $\epsilon$ we choose, more Monte Carlo samples we need to obtain a given precision. However, if $\epsilon$ is too large, the number of divisions $m$ may be huge, making the total computational resources unaffordable. This means we need some tradeoff on choosing $\epsilon$.
Actually, this is not difficult. 
By doing some trials on small systems, we can obtain some appropriate $
\epsilon$ and apply it on larger systems. With this scheme, we can calculate all values of $\beta_k$ at the beginning without any simulations, therefore different estimations of $[Z_n^{T_B}(\beta_{k-1})/Z_n^{T_B}(\beta_k)]$ can be parallelized on computers, which are further used to estimate $\ln [Z_n^{T_B}(\beta_{k-1})/Z_n^{T_B}(\beta_k)]$ and $R_n(\beta)$.

\section{Computational complexity of the annealing scheme}
\begin{figure}[htbp]\centering 
    \includegraphics[width=0.55\textwidth]{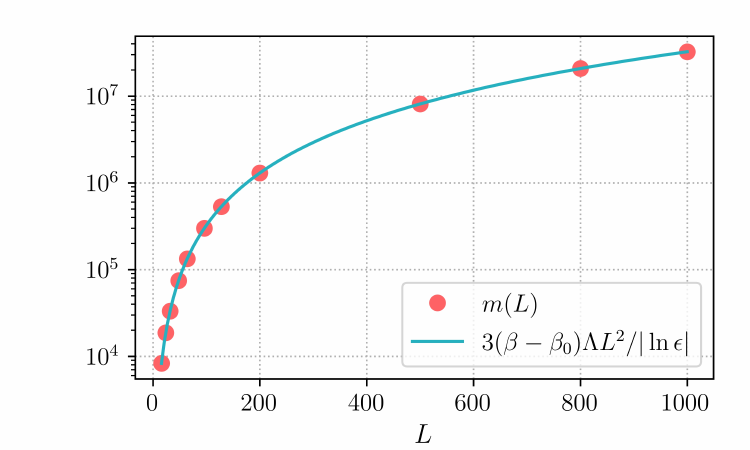}
    \caption{
        The number of subintervals $m(L)$ with different $L$ (marked in red circles) calculated with the annealing scheme we introduced. It is perfectly in accordance with the function $F(L):=\Lambda n(\beta - \beta_0)L^d/|\ln \epsilon|$, where $n=3$ for the third order of R\'enyi negativity and $d=2$ for the 2D TFIM model. For example, for $L=48$ (the largest size we simulated), we have $m(48)=74813$ and $F(48)=74804.686$.
    }\label{fig:divisions}
\end{figure}

In this section, we prove that the annealing scheme makes our algorithm have polynomial computational complexity. 

For each $\langle (\beta_{k-1}/\beta_k)^s\rangle\approx \epsilon$ close to 1 to ensure the importance sampling, the number of samples required to efficiently estimate it should not increase exponentially with the system size, provided the relative error does not diverge exponentially. 
Denote $O\equiv (\beta_{k-1}/\beta_k)^s$, then 
\begin{equation}
    \frac{\Delta O}{\langle O\rangle} = 
\frac{\sqrt{\langle O^2\rangle-\langle O\rangle^2}}{\langle O\rangle\sqrt{N}}
=
\frac{\sqrt{\langle O^2\rangle/\langle O\rangle^2-1}}{\sqrt{N}} = \frac{\sqrt{\langle O^2\rangle /\epsilon^2 - 1}}{\sqrt{N}}
\end{equation}
where $N$ is the number of samples. Remember that $O\in(0,1]$ for any possible classical configuration in the language of QMC, then $\langle O^2\rangle \le \langle O\rangle =\epsilon$, with which we finally obtain
\begin{equation}
    \frac{\Delta O}{\langle O\rangle}\le \frac{\sqrt{1/\epsilon-1}}{\sqrt{N}}
\end{equation}
which is indeed finite to any size and temperature. Therefore, as long as the annealing scheme above generates polynomial number of subintervals, the total computational complexity would be finally polynomial. 

For large $L$, we have 
\begin{equation}\label{sm:eq:complexity}
    \beta_{k-1}=\epsilon^{1/(\Lambda n\beta_kL^d)}\beta_k 
    \approx \bigg(1 + \frac{\ln \epsilon}{\Lambda n\beta_k L^d}\bigg)\beta_k
\end{equation}
which means 
\begin{equation}
    \beta_k-\beta_{k-1}\approx -\frac{\ln\epsilon}{\Lambda nL^d}
\end{equation}
thus the total number of subintervals $m=m(L)$ is around $F(L)=(\beta-\beta_0)\Lambda nL^d/|\ln\epsilon|$, which is indeed polynomial. Here we have igored the case when $\beta_k$ is small in Eq.~(\ref{sm:eq:complexity}). This actually contributes little to the value of $m(L)$ because when $\beta_k\to 0$,
$\epsilon^{1/(\Lambda n\beta_k L^d)}\to 0$ (remember that $\epsilon<1$), making the distribution of $\beta_k$ much more sparser than that of the region where $\beta_k\gg 0$.

In practical simulations on computers, we cannot really take $\beta_0=0$, so we set a sufficiently small number such as $\beta_0=10^{-8}$ used in our simulations. The adequacy of the choice of $\beta_0$ can be checked by considering a smaller one to see whether the result is convergent.
For references, in our simulations of the 2D TFIM model, we take $\Lambda=25h$, $\epsilon=0.1$ and $50$ bins for each $\langle (\beta_{k-1}/\beta_k)^s\rangle$ with $10^4$ Monte Carlo steps in each bin. 
For $\beta=0.362466667\approx \beta_c/3$ and $\beta_0=10^{-8}$, we calculate the number of subintervals $m(L)$ given by the annealing scheme, which is consistent with our estimation $F(L)=(\beta-\beta_0)\Lambda nL^d/|\ln\epsilon|$, shown in Fig.~(\ref{fig:divisions}).


Similar discussions can be applied to the case when $\lambda$ is a coupling strength in the Hamiltonian, and a modification is just replacing $\beta_k$ with $\lambda\beta_k$ to estimate $\bar s_{\lambda,\text{tot}}$ and $s_{\lambda,\text{tot}}$. We find this hypothetical linear relation actually works well practically and has similar performance as above. 

\section{Finite-size effect in simulating RN}
We first revisit the definition of R\'enyi negativity. In our earlier introduction of the algorithm, we considered an unnormalized density matrix $\rho=e^{-\beta H}$. Here we consider $\rho:=e^{-\beta H}/Z$ to be normalized for the convenience of disscussions, and the Rényi negativity can be expressed as
\begin{equation}\label{eq:rn_normalized}
    R_3 := -\ln 
    \frac{\text{tr}[(\rho^{T_{B}})^3]}{\text{tr}(\rho^3)}=-\ln\{\text{tr}[(\rho^{T_{B}})^3]\}+\ln [\text{tr}(\rho^3)]
\end{equation}
For a system in a pure state, i.e. $\text{tr}(\rho^n)=1$, the second term in Eq.~(\ref{eq:rn_normalized}) vanishes, simplifying the expression to
\begin{equation}
    R_3=-\ln\{\text{tr}[(\rho^{T_{B}})^3]\}=-\ln[\text{tr}(\rho_A)^3]
\end{equation}
as~\cite{Calabrese_2013_replicaTrick}
\begin{equation}\label{eq:prop}
    \text{tr}[(\rho^{T_{B}})^n]=
    \begin{cases}
    \text{tr}(\rho_A)^{n} & n\in\text{odd} \\
       [\text{tr}(\rho_A^{n/2})]^2  & n\in\text{even}
    \end{cases}
\end{equation}
where $\rho_A=\text{tr}_B(\rho)$ is the reduced density matrix of the subsystem $A$. 

The entanglement Rényi entropy is defined as
\begin{equation}
    S_n(\rho_A)=\frac{1}{1-n}\ln[\text{tr}(\rho_A^n)]
\end{equation} 
which implies that $R_3=2S_3$ when $\rho$ corresponds to a pure state. In the context of Monte Carlo simulations, the ground state is achieved by taking the limit $\beta\to\infty$ for $\rho=e^{-\beta H}/Z$ to filter out the excited states. If the system has a two-fold degeneracy in its ground state (e.g., as in the classical Ising model), we have 
\begin{equation}\label{eq:project}
    \begin{split}
        \lim_{\beta\to\infty} e^{-\beta H}=&\lim_{\beta\to\infty}\sum_i e^{-\beta E_i}\ket{\psi_i}\bra{\psi_i} \\ 
    =&\lim_{\beta\to\infty}e^{-\beta E_0}\bigg[\ket{\psi_0}\bra{\psi_0} + \ket{\psi_0'}\bra{\psi_0'} + \sum_{i\ne 0}e^{-\beta (E_i-E_0)}\ket{\psi_i}\bra{\psi_i}\bigg] \\ 
    =&\lim_{\beta\to\infty}e^{-\beta E_0}\bigg[\ket{\psi_0}\bra{\psi_0} + \ket{\psi_0'}\bra{\psi_0'} \bigg]
    \end{split}
\end{equation}
where $\{\ket{\psi_i}\}$ are the eigenstates of the Hamiltonian with $\ket{\psi_0}$ and $\ket{\psi_0'}$ to be the two degenerate ground states. This means that the ground state $\rho$ in QMC is basically a mixed state when the system has some degeneracy. 

In QMC simulations, the ground state of a Hamiltonian is typically achieved by taking $\beta\propto L^z$, where $z$ is the dynamical critical exponent at the critical point. This approach ensures extrapolation to the true ground state in the thermodynamic limit $(L\to\infty)$. For all the models we consider in this work, we have $z=1$, and we set $\beta=L$ to filter out the high-energy states in our work.
For finite systems, the energy gap between the asymptotically degenerate ground states vanishes exponentially with system size, and the choice of $\beta=L$ is also sufficient. 
However, the critical point effectively becomes an extreme (asymptotically gapless) point, which shifts with increasing system size. As a result, fixing $\beta=L$ or any other linear function of $L$ may lead to a breakdown of the area law near these extreme points, as different system sizes respond differently to gaplessness at the same parameter point.

\begin{figure}[htbp]\centering 
    \includegraphics[width=0.55\textwidth]{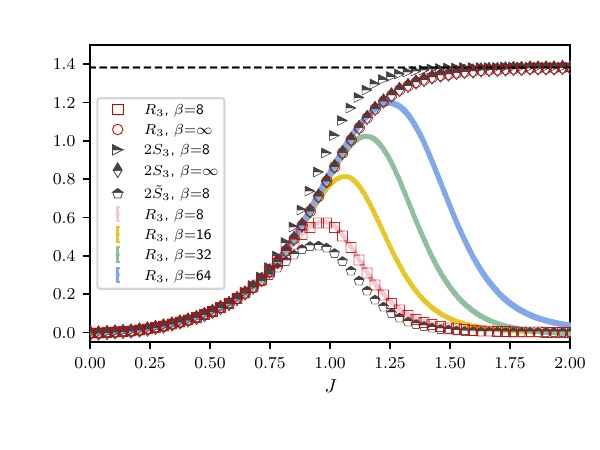}
    \caption{
        For the 1D TFIM with $h=1$, the R\'enyi negativity $R_3$ and entanglement R\'enyi entropy $S_3$ as functions of $J$, where the continuous curves are obtained from QMC simulations and the discrete markers represent results obtained from the exact diagonalization method.
    }\label{fig:fse}
\end{figure}

In closing, we discuss the difference between $R_3$ and $2S_3$ in quantifying a degenerate ground state. For illustrations, consider a simple example of 1D TFIM with $L=8$ and $h\equiv 1$. In this finite system, the true ground state remains a pure state as the two-fold degeneracy occurs at the thermodynamic limit, thus the relation $R_3(J,\beta=\infty)=S_3(J,\beta=\infty)$ holds rigorously for all values of $J$, as shown in Fig.~\ref{fig:fse}. In order to effectively filter out all high-energy states using the projector $e^{-\beta H}$ to achieve a pure state, $\beta$ must be large enough to distinguish the two lowest energy levels. 
From Fig.~\ref{fig:fse}, we observe that as $\beta$ is increased from $16$ to $64$, the behavior of $R_3(J<1,\beta)$ remains nearly unchanged, while the behavior of $R_3(J>1,\beta)$ continues to approach $R_3(J,\beta=\infty)$. Actually, this requirement is for computing $S_3$, which is only an entanglement monotone for pure states. For $R_3$, there is no need to thoroughly distinguish the two asymptotically degenerate states as we will do the extrapolation with finite systems. 
In the thermodynamic limit, the emergence of the two-fold degeneracy leads to $\text{tr}(\rho^3)=1/4$, or equivalently, $\ln [\text{tr}(\rho^3)]=-2\ln2\approx -1.386$. 
This precisely cancels the area-law correction associated with degeneracy in the entanglement entropy of a pure state~\cite{laflorencie2016quantum}, and $R_3=2S_3-2\ln 2$. When $L\to\infty$, we will have $R_3(J\ll J_c,\beta\to\infty)=R_3(J\gg J_c,\beta\to\infty)=0$ for $\rho=\lim_{\beta\to\infty}e^{-\beta H}/Z$. 
At the critical point, if the phase transition is driven by large quantum fluctuations and entanglement plays an important role at the critical point, we must observe that $R_3$ peaks at the critical point. This has been observed in our QMC simulations with $\beta=L$ for the computed system sizes. Therefore, $\beta\sim L$ is still a good choice for simulations.

\end{document}